\documentclass[letterpaper,12pt]{article}   %% LaTeX 2e (preferred)
\usepackage{osajnl2} %% do not use with REVTeX4
\usepackage[draft]{hyperref} %% optional
\usepackage{amsmath}
\begin{document}

\title{Rayleigh-Brillouin scattering profiles of air at different temperatures and pressures}

%% For REVTeX it is possible to automate superscript and e-mail callouts with the superscriptaddress option; see REVTeX4 documentation.

\author{Ziyu Gu$^1$, Benjamin Witschas$^2$, Willem van de Water$^3$, Wim Ubachs$^{1,*}$}
\address{$^1$LaserLaB, Department of Physics and Astronomy, VU University Amsterdam,
De Boelelaan 1081, 1081 HV Amsterdam, The Netherlands}

\address{$^2$Deutsches Zentrum f$\ddot{u}$r Luft- und Raumfahrt (DLR), Institut f$\ddot{u}$r Physik der Atmosph$\ddot{a}$re, Oberpfaffenhofen 82234, Germany}

\address{$^2$Department of Physics, Eindhoven University of Technology, P.O. Box 513, 5600 MB Eindhoven,
The Netherlands}
\address{$^*$Corresponding author: w.m.g.ubachs@vu.nl}

\begin{abstract}
Rayleigh-Brillouin (RB) scattering profiles for air have been recorded for the temperature range from 255 K to 340 K and the pressure range from 640 mbar to 3300 mbar, covering the conditions relevant for the Earth's atmosphere and for planned atmospheric LIDAR missions. The measurements performed at a wavelength of $\lambda= 366.8$ nm detect spontaneous RB-scattering at a $90^\circ$ scattering angle from a sensitive intra-cavity setup, delivering scattering profiles at a 1\% rms noise level or better. The experimental results have been compared to a kinetic line-shape model, the acclaimed Tenti S6 model, considered to be most appropriate for such conditions, under the assumption that air can be treated as an effective single-component gas with temperature-scaled values for the relevant macroscopic transport coefficients. The elusive transport coefficient, the bulk viscosity $\eta_b$, is effectively derived by a comparing the measurements to the model, yielding an increased trend from 1.0 to 2.5$\times10^{-5}$ kg$\cdot$m$^{-1}$$\cdot$s$^{-1}$ for the temperature interval. The calculated (Tenti S6) line shapes are consistent with experimental data at the level of 2\%, meeting the requirements for the future RB-scattering LIDAR missions in the Earth's atmosphere. However, the systematic 2\% deviation may imply that the model has a limit to describe the finest details of RB scattering in air. Finally, it is demonstrated that the RB scattering data in combination with the Tenti S6 model can be used to retrieve the actual gas temperatures.
\end{abstract}

% REPLACE WITH CORRECT OCIS CODES FOR YOUR ARTICLE
                          % NOTE: \ocis{} IS ALIASED TO \pacs{} BUT MUST
                          % FORMAT THE TERMS CORRECTLY FOR EACH JOURNAL

\maketitle %% null function with osajnl.sty

\newpage

\section{Introduction}
Light scattering in gases can be described in terms of a wavelength-dependent cross section and a scattering profile. After Rayleigh's celebrated derivation from electromagnetism the cross section, exhibiting its characteristic $\lambda^{-4}$ behavior, was understood in terms of the index of refraction of the gas~\cite{Strutt1899}. In recent years, laser techniques have made it possible to directly measure the total cross section, also accommodating small deviations from Rayleigh's formula associated with depolarization effects~\cite{Naus2000,Sneep2005}. The scattering profile function is, in the Knudsen approximation of non-colliding particles, governed by the Doppler effect adopting a purely Gaussian shape. However, this approximation only holds for the lowest pressures, while under realistic atmospheric pressures collisions and acoustic modes cannot be neglected\cite{Witschas2010}.
In more general terms, the RB-scattering profile is dependent on a dimensionless parameter $y$, which is the ratio of scattering wavelength $2\pi/k$ to the mean free path of the molecules between collisions, hence
\begin{equation}\label{equ:y parameter}
y=\frac{p}{k v_0\eta}=\frac{Nk_BT}{kv_0\eta},
\end{equation}
where $p$ is the pressure, $k$ the absolute value of the scattering wave vector $\boldsymbol{k}=\boldsymbol{k_s}-\boldsymbol{k_i}$ with $\boldsymbol{k_i}$ and $\boldsymbol{k_s}$ the wave vector of the incident and scattered light beam, $N$ the number density, $k_B$ the Boltzmann constant, $T$ the temperature, $v_0=(2k_BT/M)^{1/2}$ the thermal velocity with $M$ being molecular mass, and $\eta$ the shear viscosity.

While in the Knudsen regime $y\ll 1$ holds, in the opposite hydrodynamic regime with $y\gg 1$ the scattering profile can be represented by three distinct features with two Brillouin side-peaks accompanying the central Rayleigh peak. These Brillouin side peaks are shifted toward lower (Stokes component) or higher frequencies (anti-Stokes component) by an acoustic wave vector $\vert\boldsymbol{k_a}\vert$=$\vert\boldsymbol{k}\vert$=2$\vert\boldsymbol{k_i}\vert\sin(\theta/2)$,  $\theta$ representing the scattering angle. This implies that the shift of the Brillouin side-peaks in the frequency domain $f_b$, given by
\begin{equation}\label{equ:Brillouin shift}
f_b = 2nf\frac{v}{c}\sin(\frac{\theta}{2}),
\end{equation}
with $n$ the index of refraction, $c$ the velocity of light in vacuum, $v$ the velocity of sound, and $f$ the frequency of the incident light, depends on the scattering geometry.
In the intermediate regime, $0.3 < y <3$, of relevance for practical atmospheric conditions, the mean free path between collisions is comparable to the scattering wavelength and the continuum approach breaks down. In this regime several successful kinetic models have been designed to describe the scattering profile, based on approximate solutions of the linearized the Boltzmann equation~\cite{Boley1972,Tenti1974}. In the well-known Tenti S6 model, the collision integrals are expanded in 6 basis functions, with coefficients determined by the values of the transport coefficients: shear viscosity $\eta$, bulk viscosity $\eta_b$, thermal conductivity $\kappa$, and internal specific heat capacity $c_{int}$. % Comments from Willem
This model appears to be the most accurate model to describe the RB-scattering profile~\cite{Young1983}. Not only does it describe spontaneous RB scattering, but it also covers the recently discovered coherent RB scattering phenomenon~\cite{Pan2002,Pan2004}.

With the advent of Doppler wind LIDAR (light detection and ranging) techniques to obtain the global wind profile of the Earth's atmosphere, such as the ADM-Aeolus mission of the European Space Agency (ESA)~\cite{Aeolus2008}, a renewed interest has surfaced in experimentally measuring the RB scattering profile functions of molecular gases, particularly of air, at the highest possible accuracies~\cite{Witschas2010,Vieitez2010}. In addition, a horizontal LIDAR experiment was reported to prove the Brillouin scattering effect in the atmosphere~\cite{Witschas2012}. Since the RB-profiles cannot be measured under all possible conditions (combinations of wavelengths, scattering angles, pressure, temperature, gas composition) it is of importance to convert the experimental content into theoretical line profiles to be used in satellite retrieval algorithms. Such line profiles should be tested for an as wide as possible part of the parameter space of experimental conditions.

To date, no experiments have been conducted to verify the Tenti S6 model in air for different temperatures. Here, we expand on previous work~\cite{Witschas2010,Vieitez2010} recording spontaneous RB-scattering profiles of air at an ultraviolet wavelength of $366$ nm in a temperature range from 250 K to 340 K for various pressures. For the scattering angle a choice was made for 90$^\circ$, compromising between reasonably pronounced RB side bands and a not too small free-spectral range of the Fabry-Perot analyzing instrument. The measured scattering profiles are compared with Tenti S6 model calculations, based on a code produced by Pan~\cite{Pan_Thesis}, and adapted for comparison to our experiments~\cite{Vieitez2010}.
In these calculations the bulk viscosity, quantifying the relaxation of internal molecular modes of motion due to collisions, is a parameter. Because of the absence of internal degrees of freedom it is zero for atomic gases. It is an essential frequency-dependent quantity, but most information about its numerical value comes from experiments at ultrasound frequencies~\cite{Prangsma1973}, hence in the MHz regime. Recently, studies have been carried out with the intention to derive a value for the bulk viscosity from light scattering~\cite{Xu2003}, in particular from coherent RB-scattering~\cite{Pan2004}.
Meijer~\emph{et al.}~\cite{Meijer2010} compared values for $\eta_b$ for various gases as obtained from coherent RB scattering, acoustic measurements and molecular structure calculations, showing that there are still many unknowns in the understanding of the bulk viscosity. % quantifying the resistance of a gas to rapid compression.
In the present work we follow the path of deriving optimized Tenti S6 model functions by adapting the numerical values of the bulk viscosity, $\eta_b$.

\section{Scattering profile modeling in the kinetic regime}

Rayleigh-Brillouin scattering in gases results from density perturbations $\Delta \rho$, which can be written as a sum of distinctive contributions~\cite{Boyd2008}. The entropy perturbations of the medium at constant pressure,
\begin{equation}\label{equ:entropy_perturbations}
\Delta \rho=(\frac{\partial \rho}{\partial s})_p \cdot \Delta s,
\end{equation}
result in the central Rayleigh scattering peak, while the pressure perturbations at constant entropy,
\begin{equation}\label{equ:pressure_perturbations}
\Delta \rho=(\frac{\partial \rho}{\partial p})_s \cdot \Delta p,
\end{equation}
can be regarded as acoustic waves traveling through the medium (gases in our case), resulting in Brillouin scattering with the Stokes and anti-Stokes scattering peaks shifted by the frequency of the acoustic waves. %Since the pressure perturbations arises from the collisions between the molecules, the effect of Brillouin scattering depends on the mean free path of the molecules.
Contrary to the hydrodynamic regime, where the gas density perturbations can be sufficiently described by Navier-Stokes equations, in the intermediate regime (kinetic regime) the perturbations should be solved from the Boltzmann equation. Since the collision integral of Boltzmann equation is difficult to compute, the Tenti S6 model is based the Wang-Chang-Uhlenbeck equation~\cite{Wang-Chang1964}, which is used to construct the collision integral from the transport coefficients.

To compute the scattering profiles of air at different temperatures and pressures, the Tenti S6 model requires values for three transport coefficients, shear viscosity, thermal conductivity and bulk viscosity at the specific conditions as inputs. The assumption is made that air may be treated as a single-component gas with an effective particle mass 29.0 u~\cite{Rossing2007}, and effective transport coefficients as obtained from experiment.
Shear viscosity and thermal conductivity are known to be nearly independent of pressure. For instance, an increase of pressure $p$ from 1~bar to 50~bar will only result in 10\% change of the shear viscosity~\cite{White1998}. Because in the present study, the pressure remains below 3.5 bar, pressure effects on the transport coefficients are treated as negligible. On the other hand, temperature has a significant influence on the transport coefficients.
Values of shear viscosity $\eta$ and thermal conductivity $\kappa$ for air at certain temperatures can be calculated by~\cite{Rossing2007}:
\begin{equation}\label{equ:shear_viscosity}
\eta = \eta_0 \cdot (\frac{T}{T_0})^{3/2} \cdot \frac{T_0+T_{\eta}}{T+T_{\eta}},
\end{equation}
and
\begin{equation}\label{equ:thermal_conductivity}
\kappa = \kappa_0 \cdot (\frac{T}{T_0})^{3/2} \cdot \frac{T_0+T_A \cdot e^{-T_B/T_0}}{T+T_A \cdot e^{-T_B/T}},
\end{equation}
where $\eta_0 = 1.864\times$10$^{-5}$ kg$\cdot$m$^{-1}\cdot$s$^{-1}$  is the reference shear viscosity and $\kappa_0 = 2.624\times$10$^{-2}$ W$\cdot$K$^{-1}\cdot$m$^{-1}$ is the reference thermal conductivity, at reference temperature $T_0$ = 300 K; $T_{\eta}$ = 110.4 K, $T_A$ = 245.4 K and $T_B$ = 27.6 K are characteristic constants for air. %accuracy?.

The bulk viscosity, $\eta_b$, expressing the resistance of a gas to rapid compression,
is a parameter which is not well-understood. This parameter is effectively a second macroscopic viscosity parameter depending on the internal degrees of freedom in the molecular constituents, and therefore does not play a role ($\eta_b=0$) in the thermodynamics of mono-atomic gases~\cite{Tisza1942,Graves1999}. Here, for the measurements on air it must be considered what degrees of freedom effectively contribute to the bulk viscosity. For light-scattering experiments hypersound acoustic frequencies in the GHz range are of relevance; the acoustic frequency $f_b$ corresponds to GHz frequencies. Under room temperature conditions the vibrational degree of freedom for most gases is frozen due to its long relaxation time, and can therefore be safely neglected. The dependence on the accessible degrees of freedom causes $\eta_b$ to be temperature dependent. The value of $\eta_b$ can in principle be measured via sound absorption and a number of studies have been performed~\cite{Prangsma1973} in a variety of gases. However, such measurements yield values for $\eta_b$ in the MHz frequency domain, and they are most likely not directly applicable to the GHz regime of hypersound as is assessed via Rayleigh-Brillouin scattering. Pan \emph{et al.}~\cite{Pan2004,Pan2005} have proposed to measure the bulk viscosity through RB-scattering experiments, in particular for the case of its coherent form. %

The RB-profile depends on the macroscopic transport coefficients, the shear viscosity $\eta$, the heat conductivity $\kappa$, the internal specific heat capacity $c_{int}$, and the bulk viscosity $\eta_b$, as well as the temperature $T$, pressure $p$ of the gas, the mass of the particle constituents, and the wavelength $\lambda$ or the frequency $f$ of the incident light beam, and the scattering angle $\theta$. While the laser and scattering parameters can be measured, $c_{int}$ and particle mass readily caculated, the transport coefficients $\eta$ and $\kappa$ are known from literature to high accuracy. Hence the final, more elusive transport coefficient $\eta_b$ can be derived from RB-scattering if a model is established that links the scattering profile to the transport coefficients. By this means Pan \emph{et al.} found large discrepancies, up to orders of magnitude, between values for $\eta_b$ as measured by light scattering compared to previous measurements via sound absorption~\cite{Pan2004,Pan2005} for the specific example of CO$_2$. In this work, we determine the effective bulk viscosity as the value which provides the best fit between the measured line profiles (at high pressures) and the one computed from the Tenti S6 model. It remains a question whether this approach is adequate. As a test, the temperature dependence of the bulk viscosity will be determined, % Willem's comment 1.4.2
and the obtained values will be further verified with low pressure data, assuming the pressure dependence of $\eta_b$ is negligible (similar to the shear viscosity).

\section{Experimental Setup}
Details of the experimental setup and methods for measuring RB scattering profiles have been reported in~\cite{Gu2012}. A narrow bandwidth frequency-doubled titanium:sapphire laser delivers a collimated beam of 500 mW of continuous wave ultraviolet light at 366.8 nm. This intensity is further amplified by an order of magnitude in an enhancement cavity. In a beam focus inside the enhancement cavity, a gas scattering cell is mounted, designed to permit a controlled and stable temperature setting between 250 K and 340 K and a pressure setting between 0 bar and 4 bar. Rayleigh-Brillouin scattered photons are collected at a scattering angle of 90$^{\circ}$ and subsequently analyzed spectrally by a home-built plano-concave Fabry-Perot interferometer (FPI) with an instrument linewidth of 232 MHz, and an effective free spectral range (FSR) of 7440 MHz.  A high gain photo-multiplier tube (PMT) is used for detection and to record the scattered light passing through the FPI. A typical recording period is around 3 hours, during which a typical frequency span of 400 GHz (corresponding to some 50 FSR) is covered. Procedures are followed that correct for drift of the laser frequency and the FPI during data recording, %
and then all data collected in a 3 hour scan are averaged and normalized to area unity. %$\int_{-{FSR}/2}^{{FSR}/2}\,I(f)\,df=1$
The scattering profiles are finally compared with the numerical calculations, performed for the exact measurement conditions, and convolved with the instrument function of the FPI (referred to as convolved Tenti S6 model afterwards).

Although the dark counts of the PMT, and the background of the Airy function corresponding to the overlap of consecutive FSRs, have already been taken into account in the calculations, the background of the measurements is always higher than the background of the calculations. This phenomenon had been addressed previously~\cite{Vieitez2010} to broadband fluorescence of the cell windows. However, fluorescence is unlikely to play a role here, because non-coated windows are used for the laser beam to pass through the cell and bare fused silica exhibits a fluorescence spectrum longward of 400 nm~\cite{Fluorescence}, while this part of the spectrum is filtered before detection. Raman scattering, amounting to $\sim$2.5\% of the total cross section, is another possible source of background. The rotational Raman scattered light, with a large number of individual components of width $\sim 3$ GHz distributed over several nm, is effectively spread over many modes of the FPI, resulting in a broad structureless background.
This additional background, which amounts to $\sim 2$\% of the central Rayleigh peak intensity, is corrected by using the same method as in~\cite{Vieitez2010}. In addition stray light from the cell and optics might play a role. But this would result in a narrow frequency window at the central frequency (see below).

The sample gas, air, is cooled to $-$40$^{\circ}$C to freeze out the water content to 128 ppm, and then reheated before using. While charging the cell, particles larger than 500 nm were removed by an aerosol filter in the gas inlet line.
For each measurement, the gas scattering cell is charged to a designated pressure first and sealed at room temperature. The uncertainty of the pressure meter is calibrated to be 0.5\% of the reading. The temperature of the cell together with the gas inside is varied and kept constant by four Peltier elements and a temperature-controlled water cooling system, and simultaneously measured by two Pt-100 elements, leading to 0.5 K uncertainty. The actual pressure of each measurement is thus different from the initial pressure and calculated according to the ideal gas law, while the number density of the gas molecules in the scattering volume is the same. Therefore, the measurements are separated into 3 measurement sets by the number density in Tab.~\ref{Tab:transport_coefficients}, with the actual $p$-$T$ conditions listed.

\begin{table}[h]
{\bf \caption{\label{Tab:transport_coefficients} Conditions and values of transport coefficients for the Rayleigh-Brillouin scattering measurements. Values of $\eta$ and $\kappa$ are calculated by Eq.~(\ref{equ:shear_viscosity}) and Eq.~(\ref{equ:thermal_conductivity}), from Ref.~\cite{Rossing2007}. Values of $\eta_b$ for data Set III are obtained directly from a least-squared fit, while those for data Set I and II are calculated by Eq.~(\ref{equ:linear_fit}). The $y$ parameter for each measurement is indicated in the last column.}}\begin{center}
\begin{tabular}{c c c c c c c}

\hline
 & p & T & $\eta$ & $\eta_b$ & $\kappa$ & $y$\\
& (mbar)& (K) & (10$^{-5}$ kg$\cdot$m$^{-1}\cdot$s$^{-1}$) & (10$^{-5}$ kg$\cdot$m$^{-1}\cdot$s$^{-1}$) & (10$^{-2}$ W$\cdot$K$^{-1}\cdot$m$^{-1}$) &\\
\hline
           &  643     &   254.8   &   1.624   &   0.97   &   2.265  &   0.427    \\%scan20110805_04_1
           &  704     &   276.7   &   1.734   &   1.34   &   2.441  &   0.418    \\%scan20110421_03
 Set I     &  726     &   297.1   &   1.832   &   1.68   &   2.601  &   0.396    \\%scan20110112_04_1
           &  777     &   317.8   &   1.929   &   2.03   &   2.761  &   0.390  \\%scan20110109_12
           &  827     &   337.3   &   2.017   &   2.36   &   2.908  &   0.382 \\%scan20110115_02
 \hline
           &  858     &   254.8   &   1.624   &   0.97   &   2.265   &   0.574    \\%scan20110804_01_2
           &  947     &   276.7   &   1.734   &   1.34   &   2.441   &   0.564  \\%scan20110421_02
Set II     &  1013    &   297.3   &   1.832   &   1.68   &   2.603   &   0.550   \\%scan20101211_01
           &  1013    &   318.3   &   1.931   &   2.04   &   2.765   &   0.505  \\%scan20101212_01_1
           &  1017    &   337.8   &   2.020   &   2.36   &   2.912   &   0.470 \\%scan20101212_04
\hline
           &  2576    &   255.0   &   1.625   &   0.96   &   2.267  &   1.718 \\%scan20110803_01
           &  2813    &   278.0   &   1.740   &   1.34   &   2.451  &   1.681 \\%scan20110512_04_1
Set III    &  2910    &   297.6   &   1.835   &   1.92   &   2.605  &   1.589  \\%scan20101214_02
           &  3128    &   319.3   &   1.936   &   1.87   &   2.773  &   1.562  \\%scan20101215_02
           &  3304    &   337.7   &   2.019   &   2.36   &   2.911  &   1.537  \\%scan20101220_01
\hline

\end{tabular}
\end{center}
%\footnotesize Parameters of air  \\ \normalsize
\end{table}

\begin{figure}[t]
  \centerline{\includegraphics[width=8.3cm]{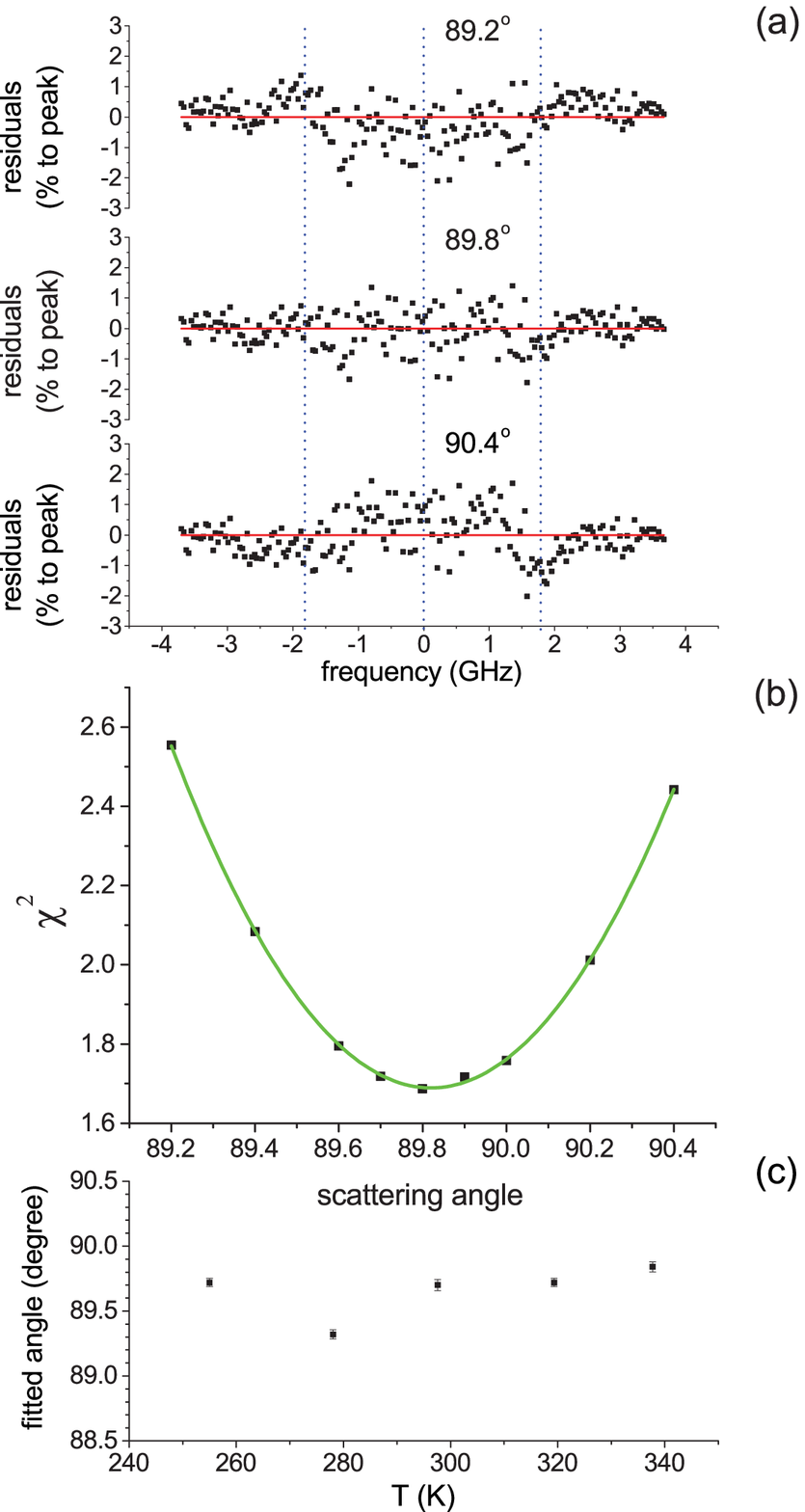}}
  \caption{\label{fig:angle_fit_with_bulk_fitted} Graphical representation of the procedure for verifying the scattering angle $\theta$; (a) Residuals between the experimental RB-scattering profile, measured for 337.7 K and 3.30 bar, and the Tenti S6 calculations, for three selected scattering angles: 89.2$^\circ$, 89.8$^\circ$ and 90.4$^\circ$; note that a value of $\eta_b=2.36 \times 10^{-5}$ kg$\cdot$m$^{-1}$$\cdot$s$^{-1}$, a result of the present study, was adopted to produce the theoretical curve. (b) The values of $\chi^2$, calculated according to the residuals, as a function of scattering angle used for Tenti S6 modeling. The green line is the parabolic fit to the $\chi^2$ values, giving a minimum at 89.8$^\circ$. The estimated error (1$\sigma$) for this angle determination is less than 0.1$^\circ$. (c) Optimized scattering angles together with their standard errors for all the measurements in data Set III). }
\end{figure}

The value of the scattering angle was previously determined via assessment of the geometrical layout of the experimental setup, with an uncertainty of 0.9$^\circ$~\cite{Vieitez2010}.
This value can be further verified from the actual scattering data, as the RB-scattering profile is rather sensitive to the scattering angle~\cite{Gu2012}. A complicating factor is that for data Set III both the bulk viscosity $\eta_b$ and the scattering angle $\theta$ influence the RB-profile, with $\eta_b$ value having a larger impact.
Moreover the scattering angle $\theta$ mainly affects the total width of the RB-scattering profile, where the bulk viscosity $\eta_b$ determines the pronounced occurrence of Brillouin side features; hence the two parameters are not strongly correlated.
For this reason, a procedure is adopted to determine an initial value for the bulk viscosity $\eta_b$, and then subsequently perform a fit to the scattering angle, minimizing the residual between model and experiment. The $\chi^2$ residual is defined as~\cite{Meijer2010}:
\begin{equation}
\label{equ:least-square fit}
\chi ^2=\frac{1}{N} \sum_{i=1}^N \frac{[I_e(f_i)-I_m(f_i)]^2}{\sigma^2(f_i)},
\end{equation}
where $I_e(f_i)$ and $I_m(f_i)$ are the experimental and modeled amplitude of the spectrum at frequency $f_i$, and $\sigma(f_i)$ the statistical (Poisson) error. An example of such a least-squares minimization to angle $\theta$, for an experimental RB scattering profile of air at 337.7 K and 3.30 bar is presented in Fig.~\ref{fig:angle_fit_with_bulk_fitted}. Panel (a) shows the residuals between the measurement (black dots) and the modeled scattering profile (red line), when three different scattering angles, 89.2$^\circ$, 89.8$^\circ$ and 90.4$^\circ$, are used for modeling. The $\chi^2$-values calculated from the residuals are 2.55, 1.68, and 2.44 for these three angles. In Fig.~\ref{fig:angle_fit_with_bulk_fitted} (b), values of $\chi^2$ are plotted as a function of scattering angles $\theta$ employed in the S6 model. The $\chi^2$-fit, represented by the full (green) line yields an optimized scattering of $\theta = 89.8^\circ$ with a $1\sigma$ standard deviation less than 0.1$^\circ$. This agrees well, with the direct geometrical assessment of the angle 90 $\pm$ 0.9$^\circ$. The determined scattering angles for all the 5 temperatures in data Set III are plotted in Fig.~\ref{fig:angle_fit_with_bulk_fitted} (c),  indicating that the scattering angles for the same temperature settings (hence number densities) are the same. It is noted that the two left most points in (c) pertain to data recorded with time intervals of several months; after such down-time, a full alignment of the optical system had to be performed, explaining the 0.4$^\circ$ scattering angle deviation.

This approach of optimizing scattering angles is further applied to the other two data sets, yielding an averaged $\theta$ to be 90.2$^\circ$ for the data Set I, $90.4^\circ$ for the data Set II, and 89.7$^\circ$ for the data Set III. Therefore, all the derived values of $\theta$ were found in the range 90 $\pm$ 0.9$^\circ$. The slight deviations are attributed to realignment of the laser beam-path through the scattering cell. The number density variation inside the cell causes the index of refraction to change and therewith the angle of the laser beam with respect to the Brewster windows; note that outside the cell atmospheric pressure is maintained~\cite{Gu2012}. In order to keep the enhancement cavity at optimized circulating intensity angular variations of a few $0.1^\circ$ have to be imposed consequently.

\section{Measurements and Analysis}
\label{sec:results}

\begin{figure}[t]
  \centerline{\includegraphics[width=15cm]{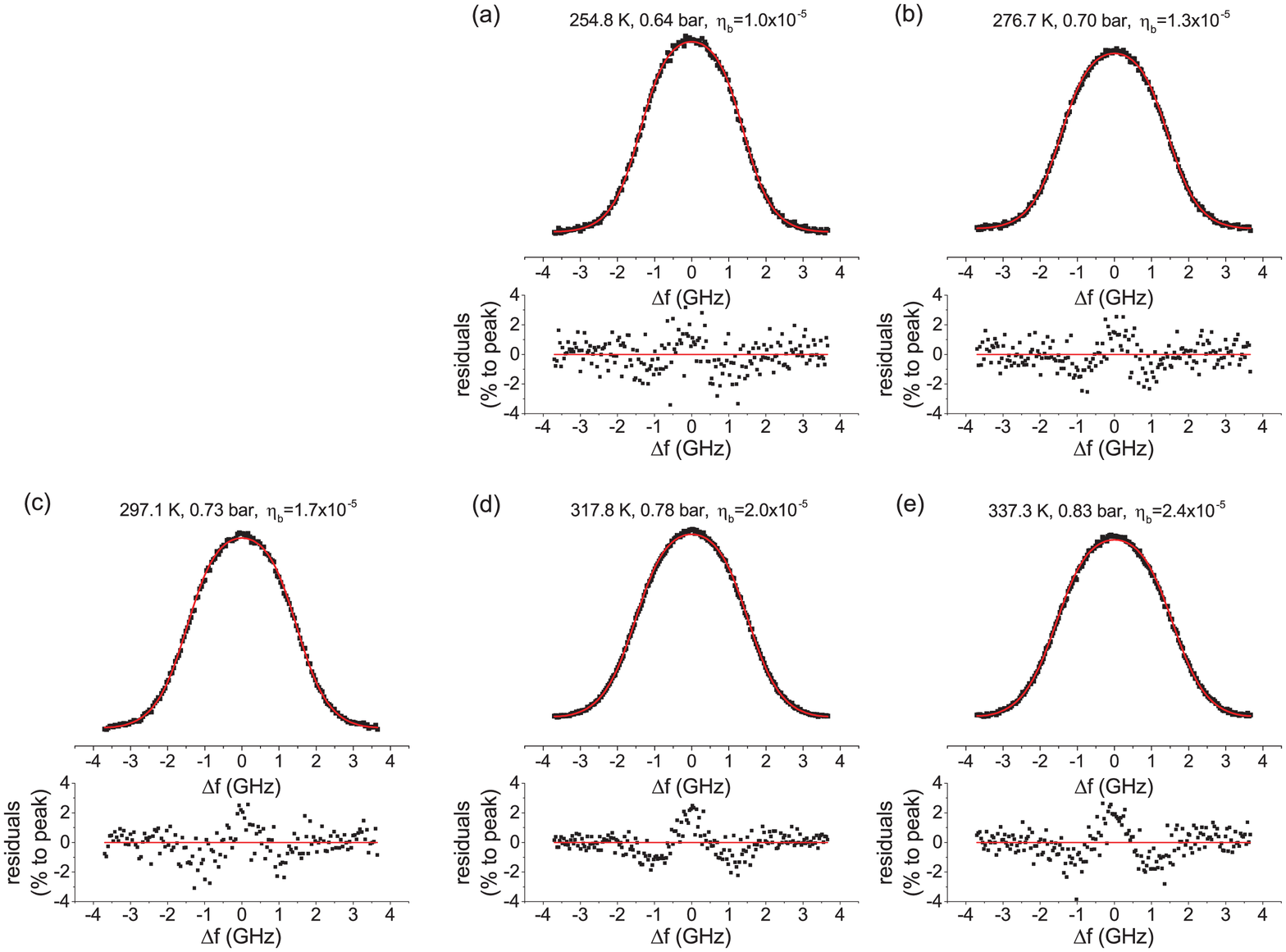}}
  \caption{\label{fig:075bar} Data set I; Normalized Rayleigh-Brillouin scattering profiles of air recorded at $\lambda=366.8$ nm, for pressures $\sim 725$ mbar and temperatures as indicated. The scattering angle for this data set was determined as $\theta = 90.2^\circ$ in the previous section. Experimental data (black dots) are compared with the convolved Tenti S6 model calculations (red line), with the input parameters listed in Table~\ref{Tab:transport_coefficients}. Values of $\eta_b$ at different temperatures are calculated from Eq.~(\ref{equ:linear_fit}).}
\end{figure}

\begin{figure}[t]
  \centerline{\includegraphics[width=15cm]{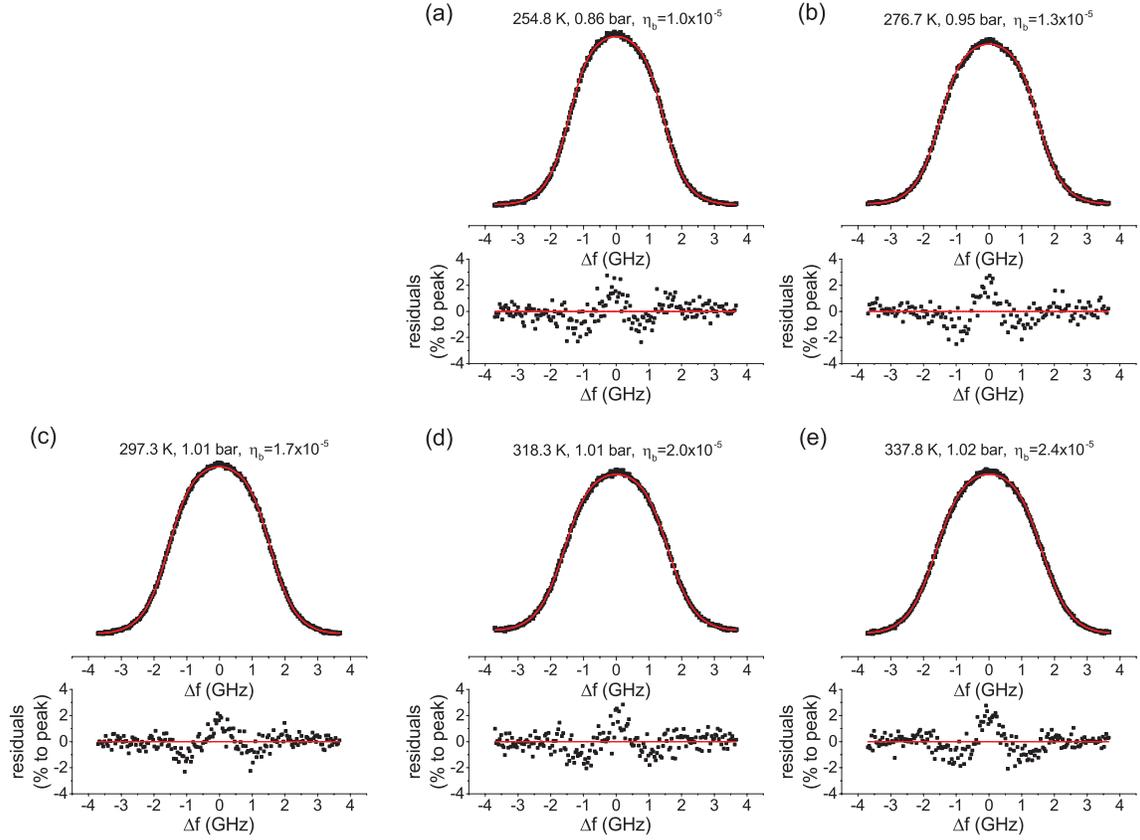}}
  \caption{\label{fig:1bar} Data set II; Normalized Rayleigh-Brillouin scattering profiles of air recorded for pressures $\sim 1000$ mbar and temperatures as indicated. The scattering angle for this data set was determined as $\theta = 90.4^\circ$ in the previous section. Values of $\eta_b$ at different temperatures are calculated from Eq.~(\ref{equ:linear_fit}).}
\end{figure}

\begin{figure}[t]
  \centerline{\includegraphics[width=15cm]{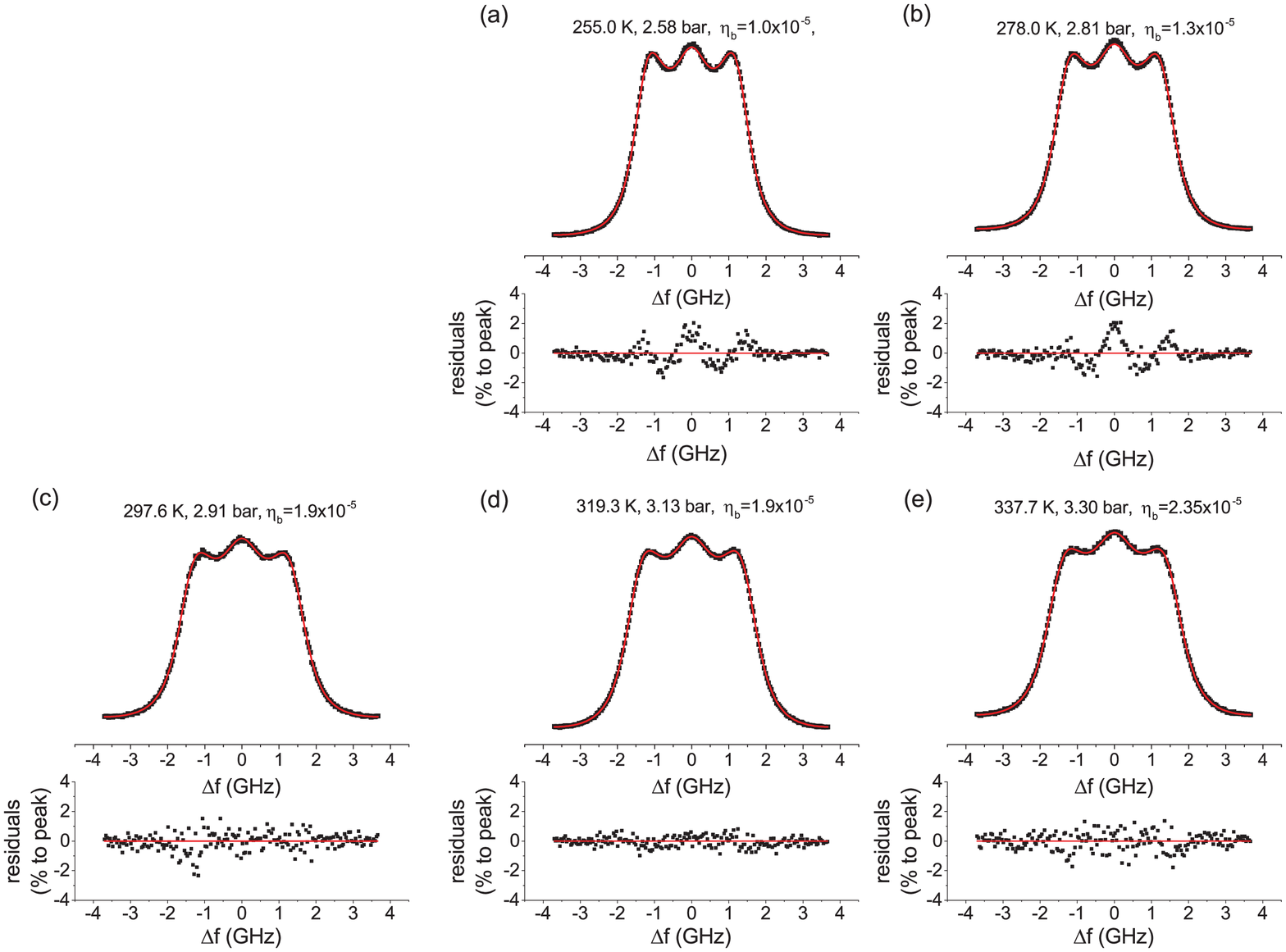}}
  \caption{\label{fig:3bar} Measurement set III; Normalized Rayleigh-Brillouin scattering profiles of air recorded for pressures $\sim 3000$ mbar. The scattering angle for this data set was determined as $\theta = 89.7^\circ$ in the previous section. Values of $\eta_b$ at different temperatures are directly obtained from the least-squared fit.}
\end{figure}

In the present study, the relevant pressure and temperature parameter space for Rayleigh-Brillouin scattering in air is mapped. A choice was made for three different initial charging pressures (725 mbar, 1000 mbar, and 3000 mbar), combined with five different temperature settings at 255 K, 277 K, 297 K, 318 K and 337 K, at intervals of $\sim$20 K. While the super-atmospheric pressure of 3 bar is not relevant in an Earth atmospheric context, its data are important for a stringent test of the Tenti S6 model, in view of the higher signal-to-noise ratio obtained and the more pronounced Brillouin side peaks occurring at higher pressures. The experimental data, plotted in black dots in Figs.~\ref{fig:075bar}, \ref{fig:1bar}, and \ref{fig:3bar}, present the central scientific content of this study. Note that the temperature settings are not exactly reproduced for different data sets; actual values of $p$ and $T$ are given in the figures and in Table~\ref{Tab:transport_coefficients}.
The RB-data were analyzed adopting values for the scattering angle $\theta$ as discussed in the experimental section.

\begin{figure}[t]
  \centerline{\includegraphics[width=8.3cm]{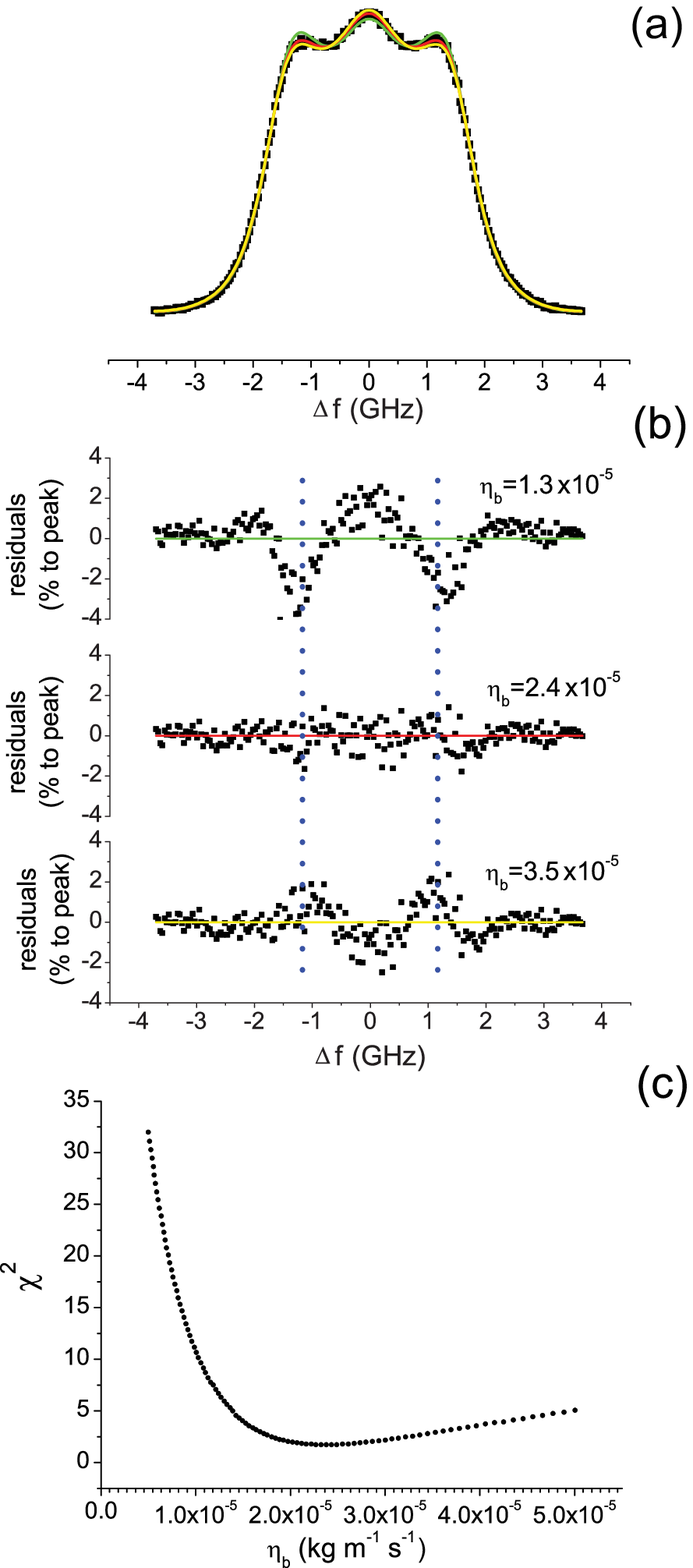}}
  \caption{\label{fig:Air_3bar_65C_eta_b} (a) Experimental Rayleigh-Brillouin scattering profile in air for 3.30 bar and 337.7 K (black dots), and convolved Tenti S6 calculations for bulk viscosity being 1.3 $ \times 10^5$ (green line), 2.4 $ \times 10^5$ (red line), and 3.5 $ \times 10^5$ (yellow line) kg$\cdot$m$^{-1}\cdot$s$^{-1}$ respectively. (b) Residuals between measured and theoretical scattering profiles for three values of the bulk viscosity. The vertical dotted lines indicate the frequency where the Brillouin side peaks occur. (c) A plot of the $\chi^2$ as a function of bulk viscosity. The optimized value of bulk viscosity is found at the minimum of $\chi^2 =1.68$. The statistical error for this fit is $0.6 \times 10^{-6}$ kg$\cdot$m$^{-1}$$\cdot$s$^{-1}$. }
\end{figure}

The present data on spontaneous RB scattering were used to assess the effect of the bulk viscosity in air.
The bulk viscosity is related to the damping of sound, thus, the Brillouin side peaks, which become more pronounced at large pressures. Therefore the data recorded at 3 bar (data set III) are used to determine $\eta_b$ in a least-square procedure.
Fig.~\ref{fig:Air_3bar_65C_eta_b} shows an example of a least-squares fit for the measurement at 337.7 K and 3.30 bar.
Beside the measurement (black dots), three typical model calculations are plotted in Fig.~\ref{fig:Air_3bar_65C_eta_b} (a), for $\eta_b$ at 1.3$\times 10^5$ kg$\cdot$m$^{-1}$$\cdot$s$^{-1}$ (green), 2.4$\times 10^5$ kg$\cdot$m$^{-1}$$\cdot$s$^{-1}$ (red), and 3.5$\times 10^5$ kg$\cdot$m$^{-1}$$\cdot$s$^{-1}$ (yellow).
Fig.~\ref{fig:Air_3bar_65C_eta_b} (b) indicates the residuals between the measurement and the three calculations.
Fig.~\ref{fig:Air_3bar_65C_eta_b} (c) shows the $\chi^2$-plot versus the bulk viscosity, and provides insight into the significance of the optimized value for $\eta_b$.
The statistical error of this determination is calculated according to~\cite{Meijer2010,Vieitez2010}:
\begin{equation}\label{equ:statistical_error}
\sigma_{\eta_b}=\left(\frac{N'}{2}\frac{d^2\chi^2}{d\eta^2_b}\bigg\arrowvert_{\tilde\eta_b}\right)^{-1/2},
\end{equation}
with $N'$ the number of the independent samples in the spectrum, and $\tilde\eta_b$ the location of the minimum of $\chi^2$. Accordingly, the statistical error in this case is $0.6 \times 10^{-6}$ kg$\cdot$m$^{-1}$$\cdot$s$^{-1}$. The uncertainty of measured temperature ($\sim$ 0.5 K) and of measured pressure (0.5\% of the reading), propagate as errors in the bulk viscosity determination, because both temperature and pressure are input parameters for the Tenti S6 model calculation. By allowing the pressure and temperature to vary at their uncertainty ranges, the derived bulk viscosity changes by $1.1 \times 10^{-6}$ kg$\cdot$m$^{-1}$$\cdot$s$^{-1}$. Therefore, the total estimated error is $1.7 \times 10^{-6}$ kg$\cdot$m$^{-1}$ $\cdot$s$^{-1}$.

\begin{figure}[t]
  \centerline{\includegraphics[width=15cm]{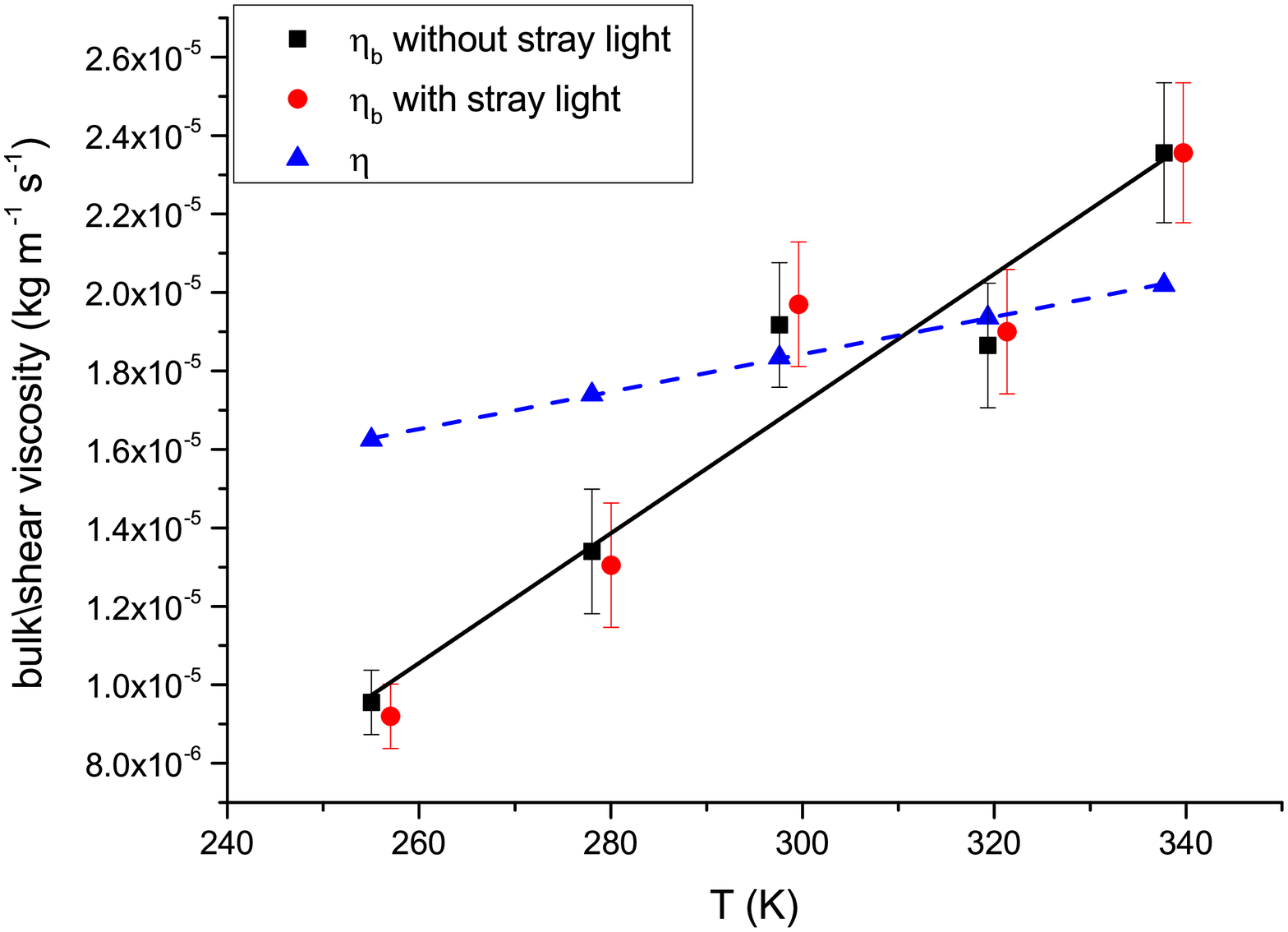}}
  \caption{\label{fig:bulk_fit_angle_fitted_shear} Bulk viscosities $\eta_b$ for air plotted as a function of temperature (black rectangular symbols) as determined from RB-scattering measurements around 3 bar air pressure and at $\lambda = 366.8$ nm. The black straight line represents a linear fit to the experimental $\eta_b$ values (see text). A comparison is made with values for the shear viscosity $\eta$ (blue upper triangles) calculated by Eq.~(\ref{equ:shear_viscosity}) in \cite{Rossing2007}. The blue dashed line is a linear fit to the $\eta$ values. The (red) circular dots represent the values of derived bulk viscosity when stray light is taken into account. Note that the red dots are offset to the right by 2 K to circumvent overlap of data points.}
\end{figure}

This fitting procedure for $\eta_b$ was applied to the other measurements in data set III.
The resulting values obtained for $\eta_b$ are plotted in (black) rectangular dots in Fig.~\ref{fig:bulk_fit_angle_fitted_shear}. In view of the observed monotonic increase a first order phenomenological model is adopted in terms of a linear dependence to which the results are fitted:
\begin{equation}\label{equ:linear_fit}
\eta_b=a+b\cdot T,
\end{equation}
This functional form represents the temperature effect on the bulk viscosity (black dashed line). This approach yields an intercept $a=(-3.33 \pm 0.60) \times10^{-5}$, and a slope $b=(1.69\pm 0.21) \times 10^{-7}$, with $\eta_b$ expressed in kg$\cdot$m$^{-1}$$\cdot$s$^{-1}$, $T$ in K, and the proportionality constants $a$ and $b$ in the corresponding units.

Finally, the convolved Tenti S6 model is calculated for all the experimental conditions pertaining to the data sets I, II, and III. A parametrization of the Tenti S6 model, as presented in~\cite{Vieitez2010} is used. Values for the scattering wavelength $\lambda=366.8$ nm, scattering angle $\theta$ as discussed above, instrument function with linewidth $\Delta\nu = 232$ MHz, particle mass 29.0 u, and values for the macroscopic transport coefficients are invoked in the model. Values for the shear viscosity and thermal conductivity are calculated according to Eqs.~(\ref{equ:shear_viscosity}) and (\ref{equ:thermal_conductivity}).
While values for the bulk viscosity for the RB-data of data set III are directly obtained from a least-squared fit to the experimental spectra, the data of sets I and II do not permit such a direct determination of bulk viscosity since at low pressures the Brillouin side peaks are less pronounced. Assuming that $\eta_b$ is independent of pressure, values for $\eta_b$ for data sets I and II are derived through the temperature dependency found in Eq.~(\ref{equ:linear_fit}).

All information on the measurement conditions and values of the transport coefficients are listed in Table~\ref{Tab:transport_coefficients}. In the last column, the values of the $y$ parameter are also indicated. The convolved Tenti S6 model calculations (red line) are compared with the measurement at each condition in Fig.~\ref{fig:075bar}, Fig.~\ref{fig:1bar} and Fig.~\ref{fig:3bar}, with the residuals in percentage of the peak amplitude of each measurement shown below. Two conclusions can be derived from the residuals: firstly, the rms noise, which is the standard deviation of the noise distribution, are at or below the 1\% level; secondly, the deviation between each measurement and the modeling calculation seems systematic but within 2\%.

Due to the small scattering cross section of the gas molecules, any RB detection setup is sensitive to stray light from cell walls, cell windows, and from the beam-steering optics. Stray light typically exhibits the same bandwidth as the incident laser beam (2 MHz), and should not be frequency shifted. Therefore, stray light will appear as a Lorentzian line located exactly in the center of the RB scattering profile with 232 MHz bandwidth, corresponding to the FWHM of the Fabry-Perot interferometer. It is noted that Mie scattering induced by aerosol particles in the scattering cell would result in a similar frequency profile; precautions were taken to avoid aerosol scattering. Because there is a systematic structure of the residuals for nearly all measurements (see Fig.~\ref{fig:075bar} to Fig.~\ref{fig:3bar}), with the measured central Rayleigh peak slightly higher than the calculation and the Brillouin peaks lower, this may be interpreted as evidence for stray light contributing to the scattering profile in the present measurements. The fact that the deviations are most apparent at the low pressure measurements, supports this hypothesis: the relative contribution of stray light should be largest at low pressures. However, it is worth noting that the additional residuals at the central Rayleigh peak always have a FWHM much larger than 232 MHz. This may imply that the increased intensity is not just attributable to stray light.

\begin{figure}
  \centerline{\includegraphics[width=15cm]{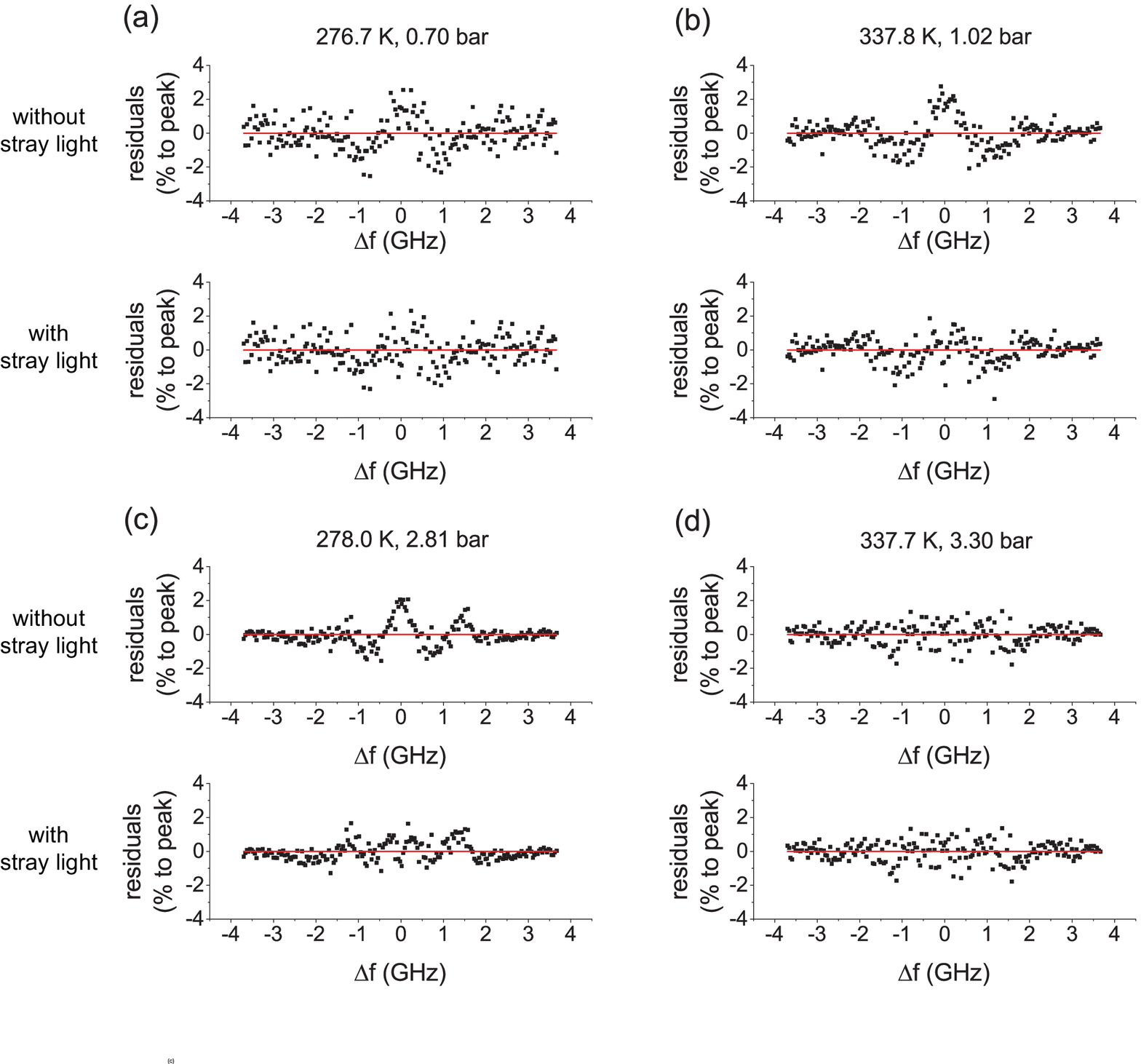}}
  \caption{\label{fig:Straylight_consideration} A comparison of residuals for four selected sample measurements of Rayleigh-Brillouin scattering profiles in air for ($p$, $T$) conditions without and with stray light included. }
\end{figure}

In order to test the stray light hypothesis, we added a spectral contribution $S\delta(f_i)$ for stray light to the modeled amplitude function $I_m(f_i)$, thus yielding a total amplitude of $I_m'(f_i)=I_m(f_i)+S\delta(f_i)$. After replacing the amplitude function $I_m(f_i)$ by $I_m'(f_i)$ in the analysis, the entire modeling procedure was repeated using the least-squares procedure of Eq.~(\ref{equ:least-square fit}). Via this means, in most cases, an improved fit to the scattering profile was obtained, for stray light intensities of $S= 0 - 0.4$ \%.
Fig.~\ref{fig:Straylight_consideration} shows four residuals of profiles with and without stray light being included.
Some cases, Fig.~\ref{fig:Straylight_consideration} (a) and (c), were selected, where indeed deviations were obviously present, and better fits are obtained if a stray light contribution is added. The $\chi^2$ are reduced from 2.15 to 1.93 when $S=$0.33\% for Fig.~\ref{fig:Straylight_consideration} (a), and the $\chi^2$ decreases from 5.67 to 4.00 when $S=$ 0.44\% for Fig.~\ref{fig:Straylight_consideration} (c). However, in several exceptional cases such as Fig.~\ref{fig:Straylight_consideration} (b), although the $\chi^2$ is reduced from 2.65 to 1.80, this stray light subtraction procedure may lead to overfitting, resulting in a conspicuous dip for the residuals $I_e(f_i)-I_m'(f_i)$ at $f_i=0$, with a systematic deviation around $\pm$ 1 GHz remaining, at the positions of the Brillouin peaks. For a few cases, Fig.~\ref{fig:Straylight_consideration} (d) for example, the stray light does not play a role: a minimum of $\chi^2$ is found when $S=$ 0\%.

Although the fitted stray light contribution ($S$) is always small (less than 0.5\%), it is questionable whether this contribution may significantly influence the values of bulk viscosity. Therefore, the values of bulk viscosity are derived again with a stray light contribution included. The newly derived values are plotted as red circular dots in Fig.~\ref{fig:bulk_fit_angle_fitted_shear}. Indeed the $\eta_b$ values are changed, but by less than $0.05\times10^{-5}$ kg$\cdot$m$^{-1}$$\cdot$s$^{-1}$ for all the temperatures, much smaller than the error margins.

\section{Discussion}

The method of extracting $\eta_b$ values from RB-scattering profiles was applied by Vietez \emph{et al.}~\cite{Vieitez2010} both for coherent RBS and for spontaneous RBS. Meijer \emph{et al.}~\cite{Meijer2010} have further detailed the CRBS studies, and at the same time performed model calculations for $\eta_b$ in various gases, while they also reviewed the literature on available quantitative information on bulk viscosity in gases.

At room temperature, Prangsma \emph{et al.}~\cite{Prangsma1973} had made measurements on sound absorption in N$_2$, yielding $\eta_b=1.28 \times 10^{-5}$ kg$\cdot$m$^{-1}$$\cdot$s$^{-1}$; in their study they reviewed a survey of 14 independent sound absorption data for near-room-temperature N$_2$ finding all values in the range $(0.75 - 1.5) \times 10^{-5}$ kg$\cdot$m$^{-1}$$\cdot$s$^{-1}$. In comparison light scattering yields $\eta_b= (2.6 \pm 0.5) \times 10^{-5}$ kg$\cdot$m$^{-1}$$\cdot$s$^{-1}$ for CRBS~\cite{Meijer2010} and $\eta_b= 2.2 \times 10^{-5}$ kg$\cdot$m$^{-1}$$\cdot$s$^{-1}$ for SRBS~\cite{Vieitez2010}. Pan \emph{et al.}~\cite{Pan2004} and Cornella \emph{et al.}~\cite{Cornella2012} used a value of $\eta_b=1.28 \times 10^{-5}$ kg$\cdot$m$^{-1}$$\cdot$s$^{-1}$ for N$_2$ at room temperature in their modeling of RB-profiles. So for N$_2$ at room temperature differing values for $\eta_b$ are found for low and high frequencies, but the difference is only a factor 2 - 3, much less than in the case of CO$_2$, where orders of magnitude discrepancies were obtained~\cite{Pan2005}.

In the study by Prangsma \emph{et al.}~\cite{Prangsma1973} temperature effects on the bulk viscosity were also investigated, for N$_2$ and other singular molecule gases. Values for $\eta_b$(N$_2$) were obtained from 77 K to 300~K, yielding linearly increasing values $\eta_b= (0.2 - 1.3)\times 10^{-5}$ kg$\cdot$m$^{-1}$$\cdot$s$^{-1}$~\cite{Prangsma1973}. Recently, a similar linear $T$-dependence of bulk viscosity for N$_2$, with the values from 0.7$\times 10^{-5}$ kg$\cdot$m$^{-1}$$\cdot$s$^{-1}$ at 255 K to 2.0$\times 10^{-5}$  kg$\cdot$m$^{-1}$$\cdot$s$^{-1}$ at 337 K,  has been reported by our group~\cite{ZiyuGu2013}. To the authors' knowledge, no $T$-dependence of the bulk viscosity has been investigated for air, which is a mixture of gases.

In the present study $\eta_b$ is derived from SRBS in air for various temperatures, results of which are displayed in Fig.~\ref{fig:bulk_fit_angle_fitted_shear}. At 297.2 K the result is $\eta_b= (1.9 \pm 0.2) \times 10^{-5}$ kg$\cdot$m$^{-1}$$\cdot$s$^{-1}$. This is in reasonable agreement with the values established for the atmospheric constituents: $\eta_b$(N$_2$) $= (2.6 \pm 0.5) \times 10^{-5}$ kg$\cdot$m$^{-1}$$\cdot$s$^{-1}$ and $\eta_b$(O$_2$) $= (2.3 \pm 0.3) \times 10^{-5}$ kg$\cdot$m$^{-1}$$\cdot$s$^{-1}$~\cite{Meijer2010}. In the analysis the assumption was made to treat air in a Tenti S6 model as a single-component gas with transport coefficients as measured for air. The present outcome shows a quantitative outcome in correspondence with findings for N$_2$ and O$_2$. While the uncertainties are rather large the data nevertheless demonstrate a clear example of a bulk viscosity increasing with temperature. Assuming that all rotational degrees of freedom are accessible, the temperature dependence of $\eta_b$ owes to the collisional velocity dependence of the rotational relaxation time.  For hard--sphere collisions, this would imply $\eta_b \propto T^{1/2}$, which is not observed.  A similar temperature dependence $\eta \propto T^{1/2}$ would be expected for the shear viscosity  \cite{chapman1991mathematical}. That the assumption of hard--sphere collisions is too simple is already implicit in Eq.~(\ref{equ:shear_viscosity}) for the shear viscosity.

It has been argued that there should exist a fixed ratio between shear viscosity and bulk viscosity, independent of temperature and pressure~\cite{Shimizu1986}. This assumption, however, was questioned, because the bulk viscosity should exhibit a frequency-dependence due to the competition between the internal relaxation time of molecules and the period of sound waves~\cite{Graves1999}.
Moreover, the velocity dependence of momentum exchange collisions can be different from those for the relaxation of rotational modes.
The present results on the temperature-dependent bulk viscosities in air, in comparison with literature data on the shear viscosity (see Fig.~\ref{fig:bulk_fit_angle_fitted_shear}), show that the increase in $\eta_b$ is more rapid than for $\eta$. This was also the case for previous studies\cite{Prangsma1973,ZiyuGu2013}.

\section{Temperature retrieval from SRBS profiles }

Besides determining wind velocities, retrieval of temperatures of the atmosphere with Rayleigh-Brillouin LIDAR methods is another interesting target for aero-scientists~\cite{Shimizu1986}. However, due to the low RB-scattering cross section and complicated scattering profile calculations for air, the accuracy is limited~\cite{Young1983}. Therefore, Brillouin LIDAR techniques have only been applied to measure the temperature of water~\cite{Schorstein2009,KunLiang2013}. Here we demonstrate that with the Rayleigh-Brillouin scattering method, in comparison with the Tenti S6 model, it is possible to measure the temperature of air under atmospheric pressures.

\begin{figure}
  \centerline{\includegraphics[width=8.3cm]{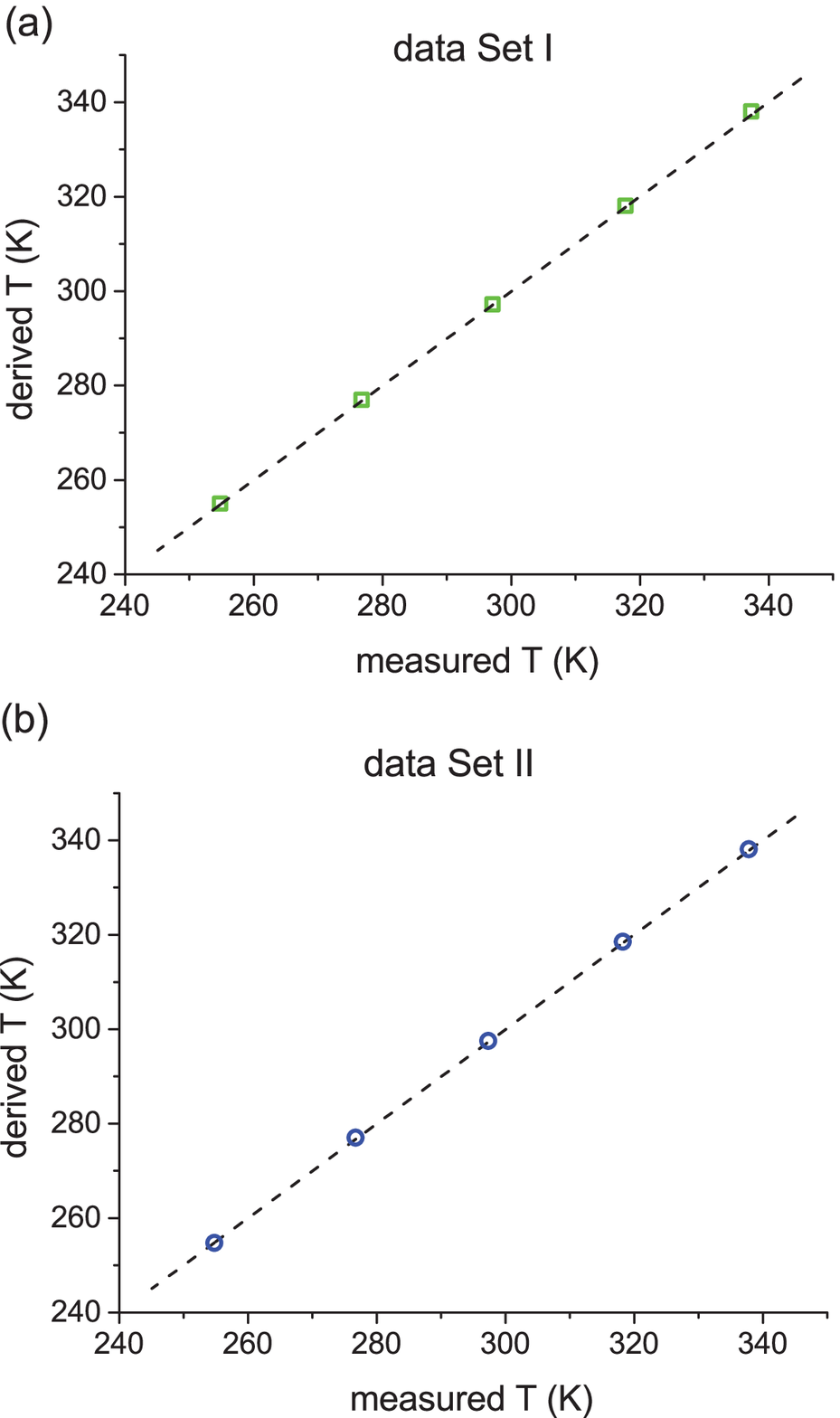}}
  \caption{\label{fig:T_derivation} Temperature retrieval form RB-scattering profiles in air. The derived temperatures for (a) data Set I and (b) data Set II, as function of measured temperatures. The dashed lines indicate where derived and measured values are equivalent.}
\end{figure}

Fig.~\ref{fig:T_derivation} shows the comparisons between the measured temperatures (with the PT-100 temperature sensors in the experimental setup) and derived temperatures for data Set I and data Set II. The derived values are obtained by fitting $T$ to a minimal $\chi^2$, when the other parameters are fixed to the values used in the Tenti S6 calculations. Note that under these pressure conditions the bulk viscosity, for which we used the simple linear temperature dependence shown in Fig.~\ref{fig:bulk_fit_angle_fitted_shear}, does not play a significant role. The solid black lines represent the condition where the derived and measured temperatures are equal. It is obvious that the derived temperatures agree well with the real (measured) temperatures for all the conditions: the maximal difference is 0.4 K.

An accuracy of less than 0.5 K can be obtained, provided that all the other conditions ($p$, $\theta$) and the transport coefficients ($\eta$, $\eta_b$, and $\kappa$) are known to a high accuracy.

\section{Conclusion}

In conclusion, Rayleigh-Brillouin scattering profiles of air at 15 different pressure-temperature combinations have been recorded. From a quantitative analysis of the data at higher pressures, values for the bulk viscosity of air are determined. Values obtained for $\eta_b$ at different temperatures provide evidence for the temperature-dependent effect of this gas transport coefficient in air: $\eta_b$ tends to increase toward higher temperatures. With these newly derived values, all experimental Rayleigh-Brillouin scattering profiles in the parameter space $p=0.6-3.3$ bar and $T=255-340$ K can be reproduced by the Tenti S6 model within 2\% deviation.
The persistence of the systematic deviations between measured spectra and the model, even after allowing for the instrumental effects (scattering angle and elastic scattering), may suggest that either the Tenti model does not adequately describe our experiment, or our method to treat air as a single component gas with effective values for its molecular mass and transport coefficients leads to deviations. Also, the additionally detected background, which may be due to rotational Raman scattering, is still not fully understood.
Nevertheless, the 2\% deviation with the Tenti S6 model for a wide parameter space bears prospect of using the model for future LIDAR missions employing RB-scattering in the Earth's atmosphere. The results of temperature retrieval suggest that the Tenti S6 model could accurately predict the temperature of air within 0.5 K if all the other experimental conditions and the transport coefficients are known.

\section*{Acknowledgments}
This work has been funded by the European Space Agency (ESA) contract ESTEC-21396/07/NL/HE-CCN-2 under the supervision of Anne Grete Straume and Olivier Le Rille. The code for computing the Tenti S6 model was obtained from Xingguo Pan.

\end{document}